\documentclass[a4paper]{llncs}

\usepackage{amsfonts}
\usepackage{fancybox}
\usepackage[hyphens]{url}
\usepackage[utf8]{inputenc}
\usepackage{xspace}
\usepackage{dcolumn}
\usepackage{rccol}
\usepackage{ifpdf}
\ifpdf
 \usepackage{epic,eepicemu}
 \usepackage[pdftex]{graphicx}
 \usepackage[pdftex,bookmarks=true]{hyperref}
\else
 \usepackage{epic,eepic}
 \usepackage[dvipdfm]{graphicx}
 \ifx\hrd\undefined
  \ifx\final\undefined\def\hrd{hypertex}\else\def\hrd{dvipdfm}\fi
 \fi
 \usepackage[\hrd,bookmarks=true]{hyperref}
\fi

\def\ourtitle{Large Formal Wikis: Issues and Solutions}

\hypersetup{
  pdftitle={\ourtitle},
  pdfauthor={Jesse Alama, Kasper Brink, Lionel Mamane, Josef Urban},
  pdfkeywords={interactive theorem proving, proof dependencies, proof analysis, large formal libraries, coq, mizar},
  colorlinks=true,
  linkcolor=black,
  citecolor=black,
  breaklinks=true,
}

\def\systemname#1{\textsf{#1}\xspace}
\def\libname#1{\textsf{#1}\xspace}

\newcommand{\mizar}{\systemname{Mizar}}
\newcommand{\btrfs}{\systemname{Btrfs}}
\newcommand{\mml}{\libname{MML}}

\newcommand{\corn}{\libname{CoRN}}
\newcommand{\git}{\systemname{Git}}

\newcommand{\isabelle}{\systemname{Isabelle}}
\newcommand{\hollight}{\systemname{HOL light}}
\newcommand{\coq}{\systemname{Coq}}
\newcommand{\tmegg}{\systemname{tmEgg}}

\newcommand{\github}{\systemname{GitHub}}

\newcommand{\gitolite}{\systemname{gitolite}}

\newcommand{\TeXmacs}{T\kern-.1667em\lower.5ex\hbox{E}\kern-.125emX\kern-.1em\lower.5ex\hbox{\textsc{m\kern-.05ema\kern-.125emc\kern-.05ems}}\xspace}

\usepackage{xcolor}

\title{\ourtitle\thanks{The final publication of this paper is
  available at www.springerlink.com}}
\author{Jesse Alama\inst{1} \and Kasper Brink\inst{3} \and Lionel Mamane\inst{2} \and Josef Urban\inst{3}\thanks{The first author was funded by the FCT project
    ``Dialogical Foundations of Semantics'' (DiFoS) in the ESF
    EuroCoRes programme LogICCC (FCT LogICCC/0001/2007).
    The third author was supported during part of the research presented here by the NWO project ``Formal Interactive Mathematical Documents: Creation and Presentation'';
    during that time he was affiliated with the ICIS, Radboud University Nijmegen.
    The fourth author was supported by the NWO project ``MathWiki a Web-based Collaborative Authoring Environment for Formal Proofs''.}}
\institute{Center for Artificial Intelligence\\
  New University of Lisbon\\
  \email{j.alama@fct.unl.pt}
  \and
  \email{lionel@mamane.lu}
  \and
  Institute for Computing and Information Sciences\\
  Radboud University Nijmegen\\
  \email{josef.urban@gmail.com}
}
\authorrunning{Alama, Mamane, and Urban}

\begin{document}

\maketitle

\begin{abstract}
  We present several steps towards large formal mathematical
  wikis. The \coq{} proof assistant together with the \corn{}
  repository are added to the pool of systems handled by the general
  wiki system described in~\cite{DBLP:conf/aisc/UrbanARG10}. A smart
  re-verification scheme for the large formal libraries in the wiki is
  suggested for \mizar/MML and \coq{}/CoRN, based on recently developed
  precise tracking of mathematical dependencies.  We propose to use
  features of state-of-the-art filesystems to allow real-time cloning
  and sandboxing of the entire libraries, allowing also to extend the
  wiki to a true multi-user collaborative area. A number of related
  issues are discussed.
\end{abstract}

\section{Overview}
This paper proposes several steps towards large formal mathematical
wikis. In Section~\ref{coq} we describe how the \coq{} proof assistant
together with the CoRN repository are added to the pool of systems
fully handled by the wiki architecture proposed
in~\cite{DBLP:conf/aisc/UrbanARG10}, i.e., allowing both web-based and
version-control-based updates of the CoRN wiki, using smart
(parallelized) verification over the whole CoRN library as a
consistency guard.  Because the task of large-scale library
refactoring is still resource-intensive, an even smarter
re-verification scheme for the large formal libraries is suggested for
\mizar/MML and \coq{}/CoRN, based on precise tracking of mathematical
dependencies that we started to develop recently for the \coq{} and \mizar{}
proof assistants, see Section~\ref{Deps}. We argue for the need of an
architecture allowing easy sandboxing and thus easy cloning of the
whole large libraries. This poses technical challenges in the
real-time wiki setting, as cloning and re-verification of large formal
libraries can be both a time and space consuming operation. An
experimental solution based on the use of modern filesystems (\btrfs{} or
ZFS in our case) is suggested
in our
setting in Section~\ref{ZFS}. Solving the problem of having many
similar sandboxes and clones despite their large sizes allows us to use the
wiki as a hosting platform for many collaborating users. We propose to
use the \gitolite{} system for this purpose, and explain the overall
architecture in Section~\ref{Gitolite}. As a corollary to the
architecture based on powerful version control systems, we get
distributed wiki synchronization almost for free. In
section~\ref{MultiWiki} we conduct an experiment synchronizing our
wikis on servers in Nijmegen and in Edmonton.  Finally we discuss a
number of issues related to the project, and draw recommendations for
existing proof assistants in Section~\ref{Issues}.

\section{Introduction: Developing Formal Math Wikis}

This paper describes a third iteration in the MathWiki
development.\footnote{The first was an experimental embedding of the
  CoRN and MML repositories inside the
  ikiwiki (\url{http://ikiwiki.info/}) system, and the second
  iteration is described in our previous
  paper~\cite{DBLP:conf/aisc/UrbanARG10}.}  An agile software
development cycle typically includes several (many) loops of
requirements analysis, prototyping, coding, and testing. A wiki for
formal mathematics is an example of a strong need for the agile
approach: It is a new kind of software taking ideas from wikis,
source-code hosting systems, version control systems, interactive
verification tools and specialized editors, and strong semantic-based
code/proof assistants. Building of formal wikis seem to significantly
interact with the development of proof assistants, and their mutual
feedback influences the development of both. For example, a
  number of changes has already been done in the last year to the
  \mizar{} XML and HTML-ization code, and to the MML verification
  scripts, to accommodate the appearing wiki functionalities. See
  below for changes and recommendations to the related \coq{} mechanisms,
  and other possibly wiki-handled proof assistants. Also, see below in
  Section~\ref{Deps} for the new wiki functions that are allowed when
  precise dependency information about the formal libraries becomes
  available for a proof assistant.  

  The previous two iterations of our wiki development were necessarily
  exploratory; our work then focused on implementing the reasonably
  recognized cornerstone features of wikis.  We used version control
  mechanisms suitable both for occasional users (using web interfaces)
  and for power users (working typically locally), and allowing also
  easy migration to future more advanced models based on the
  version-controlled repositories.  We supplied HTML presentations of
  our content, enriched in various ways to make it suitable for formal
  mathematics (e.g., linking and otherwise improved presentation of
  definitions and theorems, explicit explanation of current goals of
  the verifier, etc.)  One novel problem in the formal mathematical
  context was the need to enforce validity checks on the submitted
  content; for this, we developed a model of fast (parallelized)
  automated large-scale verification, done consistently for the
  largest formal library available.

  The previous implementations already provide valuable services to
  the proof assistant users, but we focused initially only on the \mizar{} proof assistant. While library-scale refactoring and proof checking
  is a very powerful feature of the formal wikis (differentiating them
  for example from code repositories), it is still too slow for large
  libraries to allow its unlimited use in anonymous setting.  We have
  observed that users are often too shy to edit the main official
  wiki, as their actions will be visible to the whole world and
  influencing the rest of the users. A more
  structured/hierarchical/private way of developing, together with
  mechanisms for collaboration and propagation of changes from private
  experiments to main public branches are needed. Our limited implementation
  provided real-world feedback for the next steps described
  in this paper:
\begin{itemize}
\item We add \coq{} with CoRN to the pool of managed systems.
\item We describe a smarter and faster verification modes for the wikis,
  that we started to implement within proof assistants exactly because of the feedback
  from previous wiki instances.
\item We add a more fine-grained way to edit formal mathematical texts,
  making it easier to detect limited changes (and thus avoid expensive
  re-verification).
\item We manage and control users and their rights,
  allowing the wiki to be exposed to the world in a structured way
  not limited to a trusted community of users.
\item A mechanism in which the users get their own private space is
  proposed and tested, which turns out to be reasonably cheap thanks
  to usage of advanced filesystems and its crosslinking with the
  version control model.
\item A high-level development model is suggested for the formal wiki,
  designed after a recently
  proposed model~\cite{successful-git-branching} for version-controlled software development.  We extend that model by applying different correctness policies, which helps to resolve the
  tradeoffs between correctness, incrementality, and unified
  presentation discussed in~\cite{DBLP:conf/aisc/UrbanARG10}.
\end{itemize}

One aim of our work is to try to improve the visibility and usability
of formal mathematics. The field is sorely lacking an attractive,
simple, discoverable way of working with its tools. The formal
mathematics wiki we describe here is one project designed to tackle
this problem.

\section{The Generalized Formal Wiki Architecture, and its \coq{} and CoRN Instance}
\label{coq}

One of the goals of initially developing a wiki for one system
(\mizar) was to find out how much work is needed for a particular
proof assistant so that a first-cut formal wiki could be produced. An
advantage of that approach was that as \mizar{} developers we were
capable to quickly develop the missing tools, and adjust the existing
ones. Another advantage of focusing on \mizar{} initially was that the
\mizar{} Mathematical Library (MML) is one of the largest formal
mathematical libraries available, thus forcing us to deal early on
with efficiency issues that go far beyond toy-system prototypes, and
are seen in other formal libraries to a lesser
extent.

The feasibility of the \mizar{}/MML wiki prototype suggested that our general
architecture should be reasonably adaptable to any formal proof
assistant possessing certain basic properties.  The three important
features of \mizar{} making the prototype feasible seem to be: 
batch-mode (preferably easily parallelizable)
verification; fast dependency extraction (allowing some measure of
intelligence in library re-compilation based on the changed
dependencies); and availability of tools for generating HTML
representations of formal texts.  With suitable adaptation, then, any
proof assistant with these properties can, in principle, be added to
our pool of supported systems.

It turns out that the \coq{} system, and specifically the \coq{}
Repository at Nijmegen (CoRN) formal library, satisfies these
conditions quite well, allowing to largely re-use the architecture
built for \mizar{} in a \coq{}/CoRN
wiki\footnote{\url{http://mws.cs.ru.nl/cwiki/}}.

\subsection{HTML presentation of \coq{} content with coqdoc}

We found that the
coqdoc tool, part of the standard \coq{} distribution, provides a
reasonable option for enriched HTML presentation of \coq{} articles.
With some additional work, it can be readily used for the wiki
functionalities. Note that an additional layer (called Proviola) on
top of coqdoc is being developed~\cite{DBLP:conf/aisc/TankinkGMW10},
with the goal of eventually providing better presentation and other
features for interacting with \coq{} formalization in the web
setting. As in the case of \mizar{} (and perhaps even more with
nondeclarative proofs such as those of \coq{}), much implicit
information becomes available only during proof processing, and such
information is quite useful for the readers: For example, G.~Gonthier, a 
\coq{} formalizer heading the Math Components
project,\footnote{\url{ttp://www.msr-inria.inria.fr/Projects/math-components}}
asserts that his advanced proofs are human-readable, however only in
the special environment provided by the chosen \coq{} user interface.
This obviously can be improved, both by providing better (declarative)
proof styles for \coq{} (in the spirit
of~\cite{DBLP:conf/types/Corbineau07}), and by exporting the wealth of
implicit proof information in an easily consumable form, e.g.,
similarly as \mizar{} does~\cite{DBLP:conf/mkm/Urban05}.

\begin{figure}
  \centering
  \includegraphics[scale=0.34,trim = 0mm 0mm 0mm 0mm,clip]{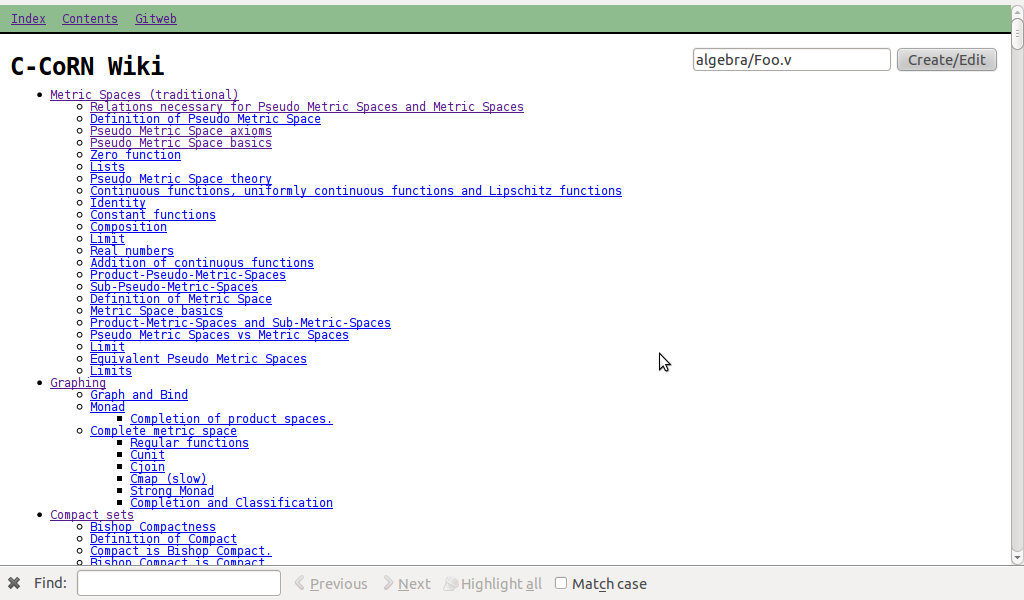}
  \caption{CoRN wiki contents page}
  \label{fig:contents}
\end{figure}

Unlike the \mizar{} HTML-ization tools (with possible exception of
  the MML Query tool~\cite{DBLP:conf/mkm/Bancerek06}), the coqdoc tool
  provides some additional functionalities like automated creation of
  indexes and tables of contents, see for example
Figure~\ref{fig:contents} for the CoRN wiki contents page.
This can be used for additional
  useful presentation of the \coq{} wiki files, and is again a motivation
  (for \mizar{} and other proof assistants) to supply such tools for
  their wikis.

\subsection{Batch-mode processing and dependency analysis with \coq}

\coq{} allows both interactive and batch-mode verification (using the
coqc tool), and also provides a special tool (coqdep) for discovering
dependencies between \coq{} files, suitable for Makefile-based
compilation and its parallelization. A difference of CoRN to MML is
that the article structure is not flat in CoRN (in \mizar{}, all
articles are just kept in one ``mml'' directory), and arbitrarily deep
directory structure has to be allowed. This poses certain challenges
when adding new files to CoRN, and taking care of their proper
compilation and HTML presentation. The current solution is that the
formal articles are really allowed to live in nested subdirectories,
while the corresponding HTML live in just one (flat) directory (this
is how the coqdoc documentation is traditionally produced), and the
correspondence between the HTML and the original article (necessary
for editing operations) is recovered by relying on the coqdoc names of
the HTML files basically containing the directory (module) structure
in them. This is a good example of a real-world library feature that
complicates the life of formal wiki developers: It would be much
easier to design a flat-structured wiki on the paper, however, if we
want to cater for real users and existing libraries, imperfect
solutions corresponding to the real world have to be used.

Interestingly, the structure of the dependencies in the CoRN
repository differs significantly from the MML. MML can really benefit
a lot from large-scale parallelization of the verification and
HTML-ization, probably because it contains many different mathematical
developments that are related only indirectly (e.g., by being based in
set theory, using some basic facts about set-theoretic functions and
relations, etc.). This is far from true for the CoRN
library. Parallelization of the CoRN verification helps comparatively
little, quite likely because the CoRN development is very
focused. Thus, even though the CoRN library is significantly smaller
than the MML (about a quarter of the size of the MML), the library
re-verification times are not significantly different when verification is
parallelized. This is a motivation for the work on finer
dependencies described in Section~\ref{Deps}.

\subsection{New CoRN development with SSReflect}
A significant issue for wiki development turns out to be the new
experimental version of CoRN, developed at Nijmegen based on the Math
Components SSReflect library. This again demonstrates some of the
real-world choices that we face as wiki developers. The first issue is
binary incompatibility. The SSReflect (Math Components) project has
introduced its own special version of the coqc binary, and standard
coqc is no longer usable with it. Obviously, providing a common wiki
for the \coq{} Standard Library and the Math Components project (even
though both are officially \coq{}-based) is thus (strictly speaking) a
fiction. One possible solution is that the compiled (.vo) files  might
still be compatible, thus allowing us to provide some clever
recompilation mechanisms for the combined libraries. The situation is
even worse with the developing version of CoRN, which relies (due to
its advanced exploration of \coq{} type
classes~\cite{DBLP:conf/itp/SpittersW10}) on both a special (fixed)
version of the coqc binary, together with a special (fixed) version of
the SSReflect library. This not only makes a joint wiki with the \coq{}
Standard Library hard to implement, but it also prevents a joint wiki with
the Math Components project (making changes to the SSReflect library,
which has to be fixed for CoRN). To handle such real issues, the
separate/private clones/branches of the wiki, used for developing
certain features and for other experiments will have to be used. This is
one of the motivations for our general proposal in Sections~\ref{ZFS}
and Section~\ref{Gitolite}.
It is noteworthy that older versions of CoRN also relied
on their own \coq{} binary, including custom ML code.
However, the features implemented by custom ML code were
partly provided by newer versions of \coq{},
and partly reimplemented in \coq{}'s LTac language.
So there is a pattern there of new developments requiring
custom \coq{} binaries which has to be taken into account when
developing real-world wikis.

\section{Using Fine-grained Dependency Information for a Large Formal Wiki}
\label{Deps}
\begin{figure}
  \centering
  \includegraphics[scale=0.34,trim = 0mm 0mm 0mm 0mm,clip]{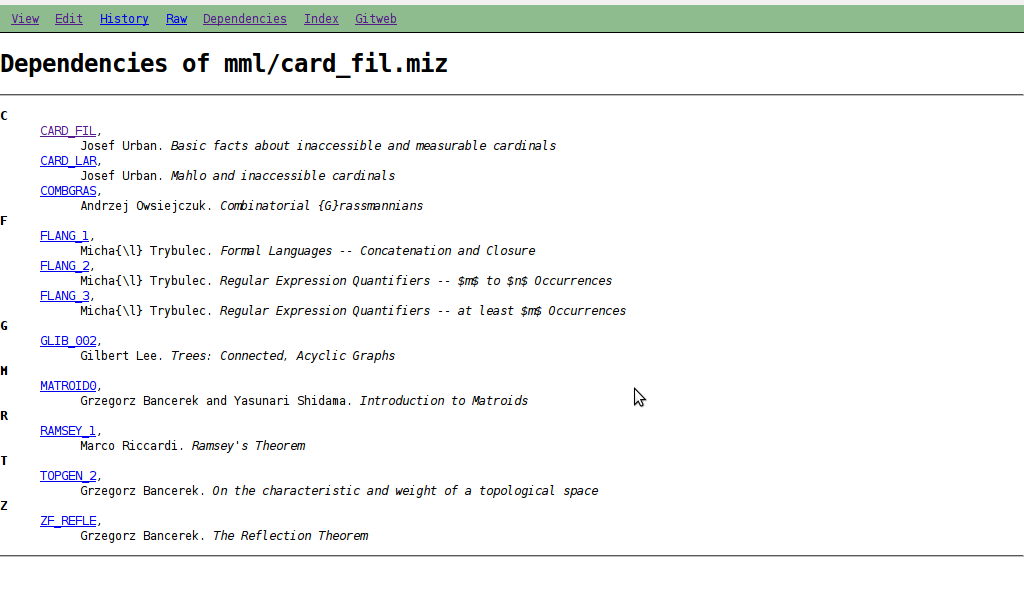}
  \caption{Aggregated fine theorem and scheme dependencies for article \texttt{CARD\_LAR}}
  \label{fig:deps1}
\end{figure}

In order to deal with the efficiency issues mentioned in previous
sections, we have started to develop tools allowing much finer
dependency tracking, and thus much finer and leaner recompilation
modes, than is currently possible with \mizar{} and \coq{}. This work is
reported in~\cite{alama-mamane-urban}. To summarize, we add a special dependency-tracking code to \coq{}, which
can now track most of the mutual dependencies of \coq{} items (theorems,
definitions, etc.), and extract the direct and transitive graph of
dependencies between these items. Similarly, but using a different
technique, we extract such fine dependencies from the \mizar{}
formalizations. For \mizar{} this is done by advanced refactoring of the
\mizar{} articles into one-item micro-articles, and computing their
minimal dependencies by a brute-force minimization algorithm. The
result of the algorithm again provides us for each item $I$ with the
precise information about which other \mizar{} items the item $I$ depends
on. This information is again compiled into graphs of direct and
indirect dependencies. The  \mizar{} wiki already allows viewing of
fine theorem and scheme dependencies aggregated for the articles, see
Figure~\ref{fig:deps1} for those of the \texttt{CARD\_LAR} article.

\subsection{Speeding up (re)verification}

It turns out that such fine dependencies have the potential to provide
significant speedups for expensive library refactorings. The following
Table~\ref{tab:stat} from~\cite{alama-mamane-urban} shows the
dependency statistics and comparison for the CoRN and MML (first 100
articles) libraries. For example, the number of direct dependency
edges computed by the fine-grained method in MML drops to 3\% in
comparison with the number of direct dependencies assumed by the
traditional coarse file-based dependencies. This is obviously a great
opportunity for the formal wiki providing very fast (and also much
more parallalizable) verification and presentation services to the
authors of formal mathematics.

\begin{table}[htb]
  \centering
  \begin{tabular}{@{\extracolsep{0.3cm}}l|*{4}{D{.}{.}{8.1}}}
    &\multicolumn{1}{c}{CoRN/item}&\multicolumn{1}{c}{CoRN/file}&\multicolumn{1}{c}{MML-100/item}&\multicolumn{1}{c}{MML-100/file}\\
    \hline
    Items& 9\:462 & 9\:462 & 9\:553 & 9\:553\\
    Deps& 175\:407 &2\:214\:396& 704\:513 & 21\:082\:287\\
    TDeps\phantom{j}& 3\:614\:445&24\:385\:358& 7\:258\:546 & 34\:974\:804\\
    P(\%)&8 & 54.5& 15.9 & 76.7 \\
    ARL&382 & 2\:577.2 & 759.8 & 3\:661.1\\
    MRL& 12.5 & 1\:183 & 155.5 & 2\:377.5\\ 
  \end{tabular}\\
  {\small
  \begin{description}
  \item[Deps] Number of dependency edges
  \item[TDeps] Number of transitive dependency edges
  \item[P] Probability that given two randomly chosen items,
    one depends (directly or indirectly) on the other, or vice-versa.
  \item[ARL] Average number of items recompiled if one item is changed.
  \item[MRL] Median number of items recompiled if one item is changed.
  \end{description}}
  \caption{Statistics of the item-based and file-based dependencies for \corn and \mml}
  \label{tab:stat}
\end{table}

\subsection{Delimited editing}

\begin{figure}
  \centering
  \includegraphics[scale=0.34,trim = 0mm 0mm 0mm 0mm,clip]{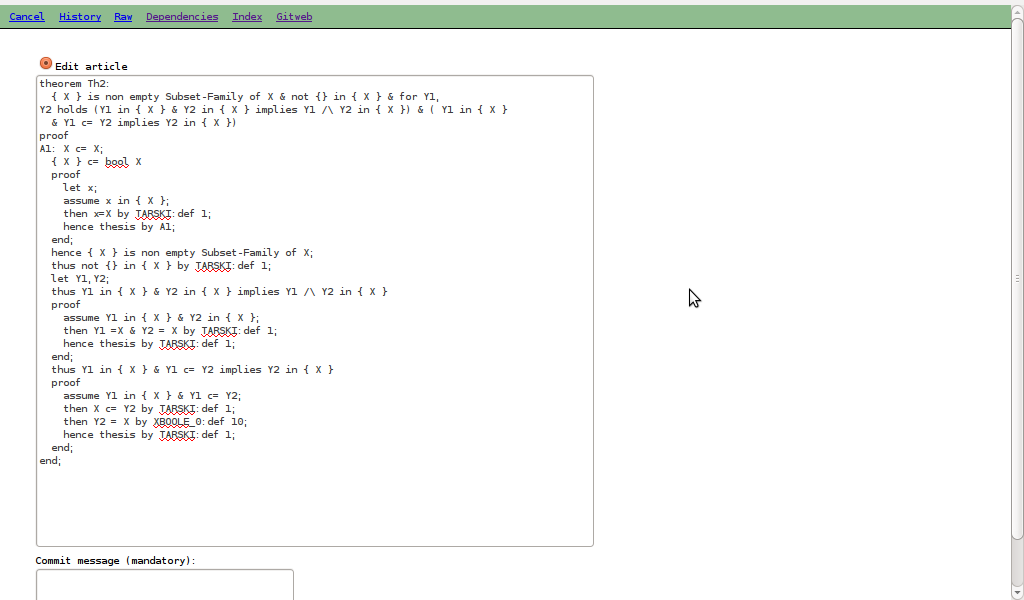}
  \caption{Delimited editing of theorem \texttt{CARD\_LAR:2}}
  \label{fig:delim-edit}
\end{figure}

The wiki now also exploits fine-grained dependency information, for the
case of \mizar, by providing delimited text editing.  The idea is to
present the user with a way to edit parts of a formal mathematical
text, rather than an entire article.  This is a formal analog of the
``Edit this section'' button in Wikipedia.  The task is to divide a
text into its constituent pieces, and provide ways of editing only
those pieces, leaving other parts intact.  The practical advantage of
such a feature is that we can be sure that edits to the text have been
made only in a small part of the text that can have only a limited
impact on other parts.  When we know that an edit is made
only to, say, the proof of a single theorem, then we do not need to
check other theorem in the text; the text as a whole is correct just
in case the new proof is correct.  If the statement of a theorem
itself is modified, it is sufficient to re-check only those other
parts of the article that explicitly use or otherwise directly depend
on this theorem. See Figure~\ref{fig:delim-edit} for an example of delimited editing of theorem \texttt{CARD\_LAR:2}.

\section{Scaling Up}
\label{ZFS}

In Section~\ref{Gitolite} we propose a wiki architecture that caters
for many users and many related developments, using the \gitolite{} tool,
and authentication policies for repository clones and branches. As
mentioned in Section~\ref{coq}, this seems to be a pressing real-world
issue, necessary for the various collaborative aspects of
formalization. Such a solution, however, forces us to deal with many
versions of the repositories, which are typically very large. The
\mizar{} HTML itself is several gigabytes in size, and in order to
be able to quickly re-compile the formal developments, we also have to
keep all intermediate compilation files around. In addition to that,
our previous implementation needed the space for at least two versions
of all these files, so that we could quickly provide a fresh sandbox
(with all the intermediate files in it) for a recompilation of only
the newly modified articles, and so that we were able to quickly return
to a clean saved state if a re-compilation in the sandbox fails. Thus,
the size of the \mizar{} wiki could reach almost 20 Gigabytes.

It is clear that with these sizes, it becomes impractical to provide a
private clone or a feature clone for hundreds (or even dozens) of
interested users. Fortunately, we can solve this by using the
\emph{copy-on-write} capabilities of modern filesystems: these mechanisms
enable us to create time- and space-efficient copies of branches in the
wiki, storing only the changes with respect to the original branch.

Currently, there are several copy-on-write filesystems under active
development; a well-known example is the ZFS filesystem, which was first
released by Sun Microsystems in 2005. Unfortunately, although ZFS is
open-source, license incompatibilities prevent it from being distributed as
part of the Linux kernel (which we use to host the MathWiki system). More
recently, work has begun on a filesystem called \btrfs{}\footnote{This stands
for ``B-tree filesystem''.},
which aims to bring many of the features of ZFS to Linux.
Included in the mainline kernel in 2009, it is not yet as stable as
traditional Linux filesystems, but its copy-on-write snapshotting is
already usable for our purposes. The functionality provided by \btrfs{} can
be combined with the architecture suggested in Section~\ref{Gitolite} to
create a system that will scale to large numbers of users and branches,
which is described below. 

The git repositories themselves are typically quite small, as they are
compressed, contain only the source files (not the intermediate and HTML
files), and additionally git allows reference sharing. Thus the main
problem are the working copies that need to be present on the server for
browsing and fast recompilation. However, these copies will typically share
a lot of content, because the users typically modify only a small part of
the large libraries, and typically start with the same main branch.

Our solution is to implement the cloning of new user repositories using
\btrfs{} snapshots. That is, we keep a working copy of the main repository in 
a separate \btrfs{} volume, and create a \emph{snapshot} (a writeable clone)
of this whenever a user clones the repository. Due to the copy-on-write
nature of \btrfs{}, this operation is efficient in terms of time and space:
creating a snapshot takes 0.03~seconds (on desktop-class hardware), and
6~KB of disk space, even for cloning very large (10G big) volumes as the
one containing the Mizar wiki. Thus, we can now
provide space for a very large number of clones and versions, and do
it practically instantaneously.

As the snapshot is modified, disk usage grows proportionally to the
size of the changes. Changing a file's metadata (e.g., updating its
last-modified-time, as required for our fast recompilation feature)
costs 10~KB on average (this is a one-time cost, paid only when the
user really makes the effort and does some acceptable changes). Modifying the
content of a file increases disk usage by the amount of newly written
data, plus a fixed overhead of about 12~KB. We have found that in
order to maximize the amount of sharing between related snapshots, it
is advisable to disable file-access-time updates on the
filesystem.\footnote{Using the ``\texttt{noatime,nodiratime}''
  filesystem options.}

Each time a repository fails to compile, and needs to be restored, we can
roll back to a previous state by discarding the latest snapshot. This is
also a fast operation, typically taking less than a second, and saving
us the necessity to maintain another 10G-large sandbox for possibly
destructive operations, and peridically using (slower) file-based
synchronization (\texttt{rsync}) with the main wiki.

The following Table~\ref{tab:cloning} documents the scalability of
\btrfs{} and its usability in our setting. It summarizes the following
experiment: The main public wiki is populated with the whole \mizar{}
library, which together with all the intermediate and HTML files takes
about 10G of an (uncompressed) \btrfs subvolume. Then we emulate 10,
100, and 200 experimental wiki clones based on the main public
wiki. Each of the clones starts as a snapshot of the main public wiki,
to which a user decides to add his new development (\mizar article)
depending on nontrivial part of the library (article
\texttt{CARD\_1}~\cite{CARD1} was used). The article is then verified
and HTML-ized, trigerring also library-wise update of various
fine-dependency indexes and HTML indexes. This process is done by
running full-scale \texttt{make} process on the whole library,
requiring reading of modification times of tens of thousands of files
in the newly created clone. Despite that, the whole process is
reasonably fast and real-time, and scales well even with hundreds
clones. The whole operation takes 6.9 seconds per clone on average for 10
clones, and 7.2 seconds on average when creating 200 clones in a
series. The average growth in overall filesystem consumption (for the
new article, its intermediate files, and updated indeces) is 5.22MB
per clone when testing with 10 clones, and 5.26MB when testing with
200 clones. To summarize, the total cost of providing 200 personalized 10G-big clones
with a newly verified article in them is only about 1GB of storage.

\begin{table}[htb]
\caption{Time and space data for 10, 100, and 200 clones
  with a new article verification}
  \centering
  \begin{tabular}{@{\extracolsep{0.5em}}r@{\hspace{1em}}|cccc}
  clones & time (s) & \multicolumn{3}{c}{disk usage (MB)}\\
  & & {\small data} & {\small metadata} & {\small total}\\
  \hline\\[-0.7em]
  10  & 6.9 & 4.71 & 0.51 & 5.22\\
  100 & 7.0 & 4.71 & 0.55 & 5.26\\
  200 & 7.2 & 4.71 & 0.55 & 5.26 
  \end{tabular}
  \label{tab:cloning}

\end{table}

\section{Many Users, Many Branches}
\label{Gitolite}

The current system now presents one version of \corn{} and the \mml{}
to the entire community. To help make the site more attractive and
useful, we would like the wiki to be a place where one can store one’s
work-in-progress; one would store one's own formal mathematical texts
and have a mechanism for interacting with other users and their
work. One could then track one’s own progress online, and possibly
follow other people’s work as well. It would be akin to a \github{}
for formal mathematics.  In this section we describe the
\git{}-based infrastructure for implementing multiple users.

The idea of extending a wiki such as ours from one anonymous user
to a secure, multiuser one, maintaining security while preserving time
and space efficiency, presents a fair number of technical
challenges. One basic question: how do we extend our \git{}-based
model?  Would we store \emph{one} repository for everyone, with
different branches for each user, or do we give each user his own
repository? How would one deal with ensuring that different users
don’t interfere with the work of other users?  How do we deal with
multiple people trying to access a repository (or repositories)?  
Note that this also leads to the problem
of storing many different (but only slightly different) copies of
large formal corpora solved in the previous section by using advanced filesystem.

\begin{figure}
  \centering
  \includegraphics[scale=0.3,trim = 0mm 90mm 0mm 0mm,clip]{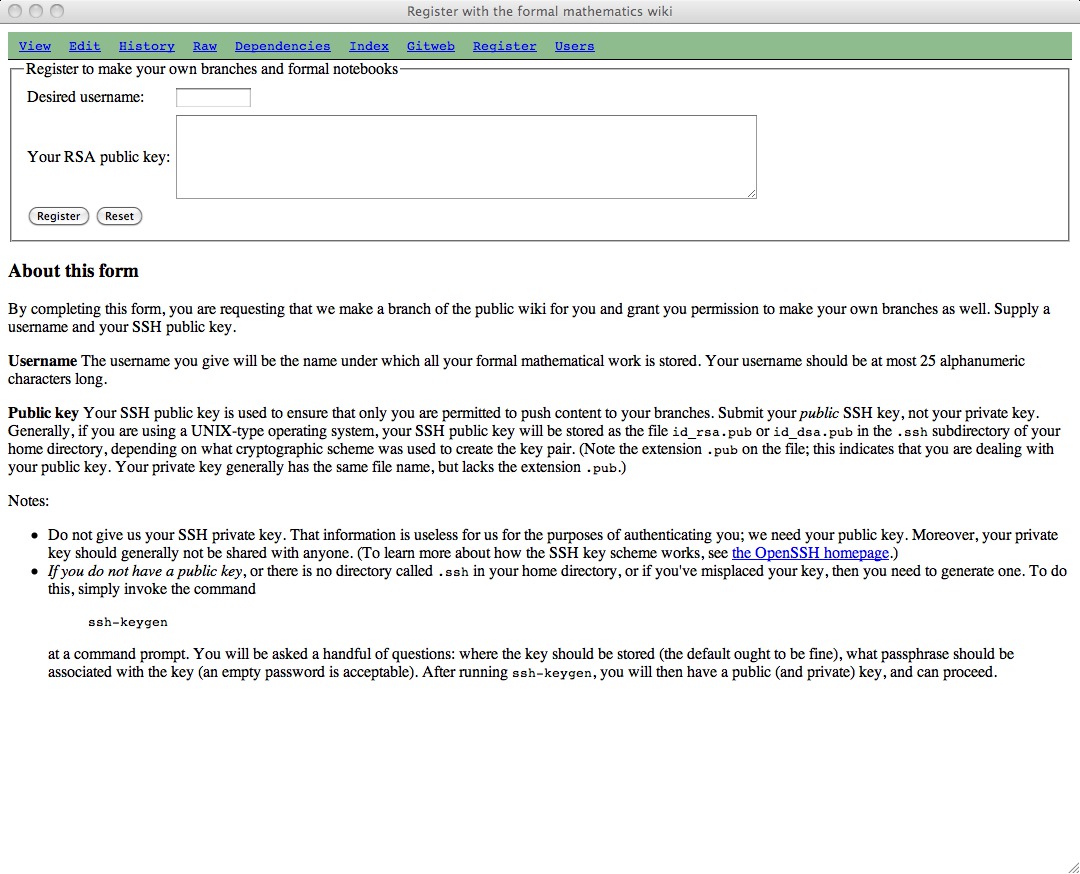}
  \caption{Registration page at our wiki}
  \label{fig:registration-page}
\end{figure}

For managing multiple users we opted for a solution based on the \gitolite{}
system.\footnote{\url{https://github.com/sitaramc/gitolite/wiki/}}
\gitolite{} adds a layer to \git{} that provides for multiple users to
access a pool of repositories, guarded by SSH keys.  With \gitolite{}
one can even set up fine-grained control over particular branches of
repositories. One can specify that certain repositories (or a
particular branch) is unavailable to a user (or group of users),
readable but not writable, or read-writable.  \gitolite{} makes
transparent use of the SSH infrastructure; once a user has provided
RSA public key to us (the registration page is shown in
Figure~\ref{fig:registration-page}), he is able to carry out these
operations via the
web page or through the traditional command-line interface to \git{}.

In addition to supporting multiple users, we also want to permit multiple
branches per user.  The following \git{} branching policy
described by V.~Driessen~\cite{successful-git-branching} 
provides a
handful of categories of branches:
\begin{quote}
  We consider origin/master to be the main branch where the source
  code of HEAD always reflects a production-ready state.  We consider
  origin/develop to be the main branch where the source code of HEAD
  always reflects a state with the latest delivered development
  changes for the next release. Some would call this the ``integration
  branch''. This is where any automatic nightly builds are built from.
\end{quote}
In addition to \emph{main} and \emph{developer} branches, we intend to
support
other kinds of branches: \emph{feature} (for work on a particular new
feature), \emph{release} (for official releases of the formal
mathematical texts), and \emph{hotfix} (fixes for critical bugs).

A \gitolite{} access implementing a Driessen-style model can be seen in Figure~\ref{fig:gitolite-policy}.
\begin{figure}[h]
  \centering
\begin{verbatim}
@all = @superusers @maintainers @developers @users @anonymous

repo    main
        RW+     =   @superusers @maintainers
        R       =   @developers @users @anonymous

repo    devel
        RW+     =   @superusers @maintainers @developers
        R       =   @users @anonymous

repo    feature/[a-zA-Z0-9].*
        C       =   @superusers @maintainers @developers
        RW+     =   @superusers @maintainers @developers
        R       =   @users @anonymous

repo    (release|hotfix)/[a-zA-Z0-9].*
        C       =   @superusers @maintainers
        RW+     =   @superusers @maintainers
        R       =   @developers @users @anonymous

repo   user/CREATOR/[a-zA-Z0-9].*
       C       =   @superusers @maintainers @developers @users 
       RW+     =   CREATOR
       R       =   @all
\end{verbatim}
  
  \caption{A \gitolite{} policy for different kinds of wiki users}
\label{fig:gitolite-policy}
\end{figure}

The intention of this policy is to divide users into certain classes
and permit certain kinds of operations (creating a branch, reading
it, reading-and-writing to it).  The user classes have the following meaning:
\begin{itemize}
\item \texttt{admin}: can do anything, has root access to the server
\item \texttt{superuser}: can do arbitrary operations on the wikis taking
  arbitrary times, can update binaries, etc
\item \texttt{maintainer}: can update the main stable wiki, start/close the
  release and hotfix branches
\item \texttt{developer}: can update the develop clone, start/close feature
  branches,
\item \texttt{user}: limited to his userspace, and inexpensive operations
\item \texttt{anonymous}: limited to the anonymous user space
\end{itemize}
The name of the repository is now also an
argument to a \git{} pre-commit or pre-receive hook, which applies a
particular verification policy to the repository. For the \emph{main} and
\emph{develop} repositories the policy should require full
verifiability, while other branches should not have to, so that these
function more like work-in-progress notebooks. (Such branches present
an interesting problem of displaying, in a helpful way, possibly
incorrect formal mathematical texts).  \gitolite{} also provides a
locking mechanism for addressing the problem of concurrent reads and
writes.


With the registration form, the wiki users can now submit their
RSA public keys to the wiki system.  Doing so adds them to \gitolite's user space, so
that they can create new (frontend) \git{} repositories (e.g., by cloning some already
existing repository).  Doing so triggers the creation of a
corresponding backend repository (\gitolite{} manages directly the
frontends, while the backend is managed indirectly by us via \git{}
hooks and CGI). The backend repositories contain the full wiki
populated with the necessary intermediate files needed for fast
re-comopilation, and obviously also with the final HTML representation of the
contents, exactly as we did in the previous one-user, one-repository
version of MathWiki. The backends themselves live in a filesystem
setup described in Section~\ref{ZFS} that re-uses space using filesystem techniques as
copy-on-write.  The result is quite a scalable platform, allowing many
users, many (related) developments, different verification and
authorization policies via gitolite and git hooks, and attempting to provide as
fast verification and HTML-ization services as possible for a given
proof assistant and library.

Note however that tasks such as re-verifying a whole large library
from scratch will always be expensive
and this should be reflected to the users. Apart from the many efficiency solutions mentioned so far, we are also experimenting with the problem of queuing pending wiki
operations. 
We should allow them to
have various superficial fast modes of verification.\footnote{For \mizar{}, one could run
  only the exporter, or also the analyzer, or the full
  verifier. These ``compiler-like'' stages actually do not have to be
  repeated in \mizar{} once they were run.} Users could have their own
queues of jobs, and would be allowed to cancel
them, if they see that some other task would invalidate the need to do
the other ones. However, when committing to the \emph{devel} or \emph{main}
branches, as mentioned, full verification should always be required.







\section{Multiple Wiki Servers and their Synchronization}\label{MultiWiki}

Mirroring is a common internet synchronization procedure used for a
number of reasons.  Mirroring increases availability by decreasing
network latency in multiple geographical locations.  Mirroring also
helps to balance network loads and supports backup of content.  An
internet mirror is \emph{live} when it is changed immediately after
its origin changes. With custom wiki software, such as
MediaWiki\footnote{\url{http://www.mediawiki.org/wiki/MediaWiki}} (the
wiki engine behind Wikipedia), there can typically be just one central
repository to which updates are made. This is no longer such a
limitation with a wiki such as our, which is built on top of a
distributed version control system.

In case of the \mizar{} part of our wiki, the practical motivation for
mirroring already exists: There are currently three reasonably
powerful servers (in Nijmegen, Edmonton, and Bialystok) where the wiki
can be installed and provide all its services. Given that
re-verification of the whole formal (e.g., \mizar{}) library is still
a costly operation, distributing the work between these servers can be
quite useful. An obvious concern is then however the desynchronization
of the developments.

This turns out to be easy to solve using the synchronization mechanism
of a distributed version control system like \git. \git{} already
comes with its own options for mirroring the changes in other
repositories, which can be easily triggered using some of its hooks
(in \git{} terminology, we are using the post-update hook on bare
repositories). Because our wiki is ``just'' a \git{} repository (with
all other functionalities implemented as appropriate hooks) that
allows pushing into it as any other \git{} repository, it turns out
that this mirroring functionality is immediately usable for live
synchronization of our wikis.  The process (for example, for two
wikis) works as follows:
\begin{itemize}
\item The wikis are initialized over the same \git{} repository.
\item A post-update hook is added to the frontend (bare) \git{}
  repository of each of the wikis, making a mirroring push
  (pushing of all new references) to the mirroring wiki's frontend
  repository.
\item Upon a successful commit/push to any of the wiki servers, the
  pushed server thus automatically updates also the mirroring wiki,
  triggering its verification and HTML-ization functions, exactly in
  the same way as a normal push to the wiki triggers these
  wiki-updating functions.
\end{itemize}

Note that this is easy with distributed version control systems such
as \git, precisely because there is no concept of a central
repository, so that all repositories are equal to each other and
implement the same functionality. It is easy also because from the
very beginning, our wiki was designed to allow arbitrary remote
pushes, not just standard wiki-like changes coming from web editing.

This mechanism also allows us to have finer mirroring policies. For
example, a realistic scenario is that each of the wiki servers by
default mirrors only changes to the main public wiki branches/clones,
and the private user branches are kept non-mirrored. This means that
the potentially costly verification operation is not duplicated on the
mirror(s) for local developments, and is done only when an important
public change is made.

\section{Conclusion and Further Issues}
\label{Issues}

We have outlined a number of steps for building on our first version
of a formal mathematics wiki.  Our aims naturally require us to make
use of several disparate technologies, including cutting-edge ones
such as smart filesystems that can cope with very large scale
datasets.

The ultimate aim of making
formal mathematics more attractive and manageable to the everyday
mathematician remains.  Extending our idea of ``research notebooks'', we would eventually like to equip our wiki with an editor with which one's
mathematical work could be carried out entirely on the web.  Collaborative tools
such as etherpad\footnote{\url{http://etherpad.org/}} are a natural
target as well.  Hooks into attractive, useful presentations of formal
proofs such as Mamane's \tmegg{} and Tankink's
Proviola~\cite{DBLP:conf/aisc/TankinkGMW10} systems can help, and
merging in powerful automation tools such as the \textsf{MizAR}
system~\cite{UrbanS10} is another obvious next step.

At the moment, our wiki supports only \mizar{} and \coq{}.  These are
but two of the actively used systems for formalized mathematics;
adding \isabelle{} and possibly \hollight{} are now within reach
thanks to our experience with \mizar{} and \coq.  Concerning \coq{},
we would like to take advantage of the ongoing Math Components
project.

Finally, we note that mappings between formal mathematics and the vast
world of ``informal'' mathematics remains rather weak.  Indeed, even
links between formal repositories is rather underdeveloped.  Linking
formal mathematical texts to some informal counterparts, such as
to Wikipedia, PlanetMath\footnote{\url{http://planetmath.org/}}, Wolfram MathWorld\footnote{\url{http://mathworld.wolfram.com}}, remains to be carried out.  For
\mizar{}, this has been achieved to some extent (providing
Wikipedia-based mapping for about two hundred MML objects), but much 
remains to be done.  It seems especially attractive, in the context of
our wiki work, to build a well-connected
corner of the World Wide Web linking formal and informal mathematics.

\bibliographystyle{splncs03}
\bibliography{mwiki2}

\end{document}